\documentclass[prd,twocolumn,showpacs,nofootinbib,preprintnumbers,floatfix]{revtex4}  
\usepackage[plainpages=false, colorlinks=true, anchorcolor=blue, linkcolor=blue, citecolor=blue, bookmarks=false]{hyperref}
\newcommand{\bqn}{\begin{eqnarray}}
\newcommand{\eqn}{\end{eqnarray}}
\usepackage{graphicx}  
\usepackage{dcolumn}   
\usepackage{bm}        
\usepackage{amssymb}   
\usepackage{natbib}
\usepackage{amsmath}   
\usepackage{longtable}  
\usepackage{float} \usepackage[utf8]{inputenc}

\DeclareUnicodeCharacter{2212}{-}
\begin{document}

\newcommand{\apjl}{Astrophys. J. Lett.}
\newcommand{\apjs}{Astrophys. J. Suppl. Ser.}
\newcommand{\aap}{Astron. \& Astrophys.}
\newcommand{\aj}{Astron. J.}
\newcommand{\pasp}{PASP}
\newcommand{\araa}{Ann. Rev. Astron. Astrophys. } 
\newcommand{\aapr}{Astronomy and Astrophysics Review}
\newcommand{\ssr}{Space Science Reviews}
\newcommand{\mnras}{Mon. Not. R. Astron. Soc.}
\newcommand{\apss} {Astrophys. and Space Science}
\newcommand{\jcap}{JCAP}
\newcommand{\na}{New Astronomy}
\newcommand{\pasj}{PASJ}
\newcommand{\pasa}{Pub. Astro. Soc. Aust.}
\newcommand{\physrep}{Physics Reports}

\title{The Hubble constant from galaxy cluster scaling-relation and SNe Ia observations: a consistency test}

\author{Kamal Bora$^{1}$}\email{ph18resch11003@iith.ac.in}

\author{R. F. L. Holanda$^{2,3,4}$}\email{holandarfl@fisica.ufrn.br}

\affiliation{$^1$ Department of Physics, Indian Institute of Technology, Hyderabad, Kandi, Telangana-502284, India }

\affiliation{$^2$Departamento de F\'{\i}sica, Universidade Federal do Rio Grande do Norte,Natal - Rio Grande do Norte, 59072-970, Brasil}

\affiliation{$^3$Departamento de F\'{\i}sica, Universidade Federal de Campina Grande, 58429-900, Campina Grande - PB, Brasil}

\affiliation{$^4$Departamento de F\'{\i}sica, Universidade Federal de Sergipe, 49100-000, Aracaju - SE, Brazil}

\begin{abstract}

In this paper, we propose a self-consistent test for a  Hubble constant estimate  using galaxy cluster and type Ia supernovae (SNe Ia) observations. The approach consists, in a first step, of obtaining the observational value of the galaxy cluster scaling-relation $Y_{SZE}D_{A}^{2}/C_{XSZ}Y_X = C $  by combining  the X-Ray and SZ observations of galaxy clusters at low redshifts ($z < 0.1$) from  the first {\it Planck mission} all-sky data set ($0.044 \leq z \leq 0.444$), along with   SNe Ia observations and making use of the cosmic distance duality relation validity. Then, by considering a flat $\Lambda$CDM model for $D_A$, the constant $C$ from the first step and the Planck prior on $\Omega_M$ parameter,  we obtain $H_0$ by using the  galaxy cluster data with $z>0.1$.  As a result, we obtain $H_0 = 73.014^{+7.435}_{-6.688}$ km/s/Mpc, which represents $9.7\%$ accuracy measurement on the Hubble constant..  We also compare our method with that one where the $C$ parameter is obtained from hydrodynamical simulations of massive galaxy clusters.

\end{abstract}
\pacs{98.80.-k, 95.35.+d, 98.80.Es}  

\maketitle

\section{Introduction}

In the last two decades, astronomical observations  have suggested that the universe behaves like a spatially flat scenario, dominated by cold dark matter (CDM) plus an exotic component endowed with large negative pressure, usually named dark energy (DE) \cite{Weinberg2013,Caldwell2009,Huterer16}.  On the other hand, one of the most important quantities to understand the cosmic history is the
current expansion rate $H_0$, its determination  has a practical and theoretical importance to many  properties of the universe (nature of dark energy, cosmic curvature, mass of neutrinos and the total number of families of relativistic particles, etc) \cite{2012arXiv1202.4459S,2021A&ARv..29....9S}. However, an accurate determination of the Hubble constant remains a puzzle in the current observational cosmology,  from which has emerged  the so-called $H_0$-tension problem: a significant tension ($5\sigma$) between  the current expansion rate of our Universe measured from the cosmic microwave background by the Planck satellite (plus a flat $\Lambda$CDM model - $H_0= 66.9 \pm 0.6$ km/s/Mpc) \cite{2020A&A...641A...6P} and from local methods (SHOES program-$H_0= 73.04 \pm 1.04$ km/s/Mpc) \cite{2021arXiv211204510R}. Curiously,  the H0LiCOW collaboration, by using  a flat $\Lambda$CDM cosmology and six gravitationally lensed quasars, found  $H_0=73.3^{+1.7}_{−1.8}$  km/s/Mpc, in 3.1$\sigma$ tension with Planck observation \cite{2020MNRAS.498.1420W}. On the other hand, by considering different cosmological probes, a $H_0$ estimate also was performed in a  model-independent way  via Gaussian process by the Ref.\cite{2021arXiv211103152G}, being found $H_0=73.78 \pm 0.84$ km/s/Mpc, which is in agreement with SH0ES and H0LiCOW estimates, but in 6.2 $\sigma$ tension with the current CMB measurements.  A detailed summary of all the measurements, tensions and related theories to resolve the $H_0$ tension can be found in ~\cite{Bethapudi,DiValentino21,2021A&ARv..29....9S,Julien,Verde,Freedman21}.
Therefore,  in order to bring some light on this puzzle, it behooves us to try additional methods to estimate $H_0$.

As a result of this $H_0$-tension, different cosmological scenarios beyond the flat $\Lambda$ cold dark matter have emerged in literature. For instance, a possibility of new physics in the form of modifying or adding energy components was discussed by  Ref.~\cite{2018JCAP...09..025M}.  Ref.~\cite{2018PhRvD..97j3529B}  advocated that the inclusion of nonlinear relativistic evolution leads to the emergence of the spatial curvature\footnote{In this context, the mean spatial curvature evolves from spatial flatness of the early universe towards slightly curved present-day universe.} alleviating the tension. A thermal production of axions coupled to heavy leptons was discussed by  Ref.\cite{2018JCAP...11..014D} as a way  to  alleviate the $H_0$ tension. Moreover, possible observational, statistical and astrophysical biases  have also been explored in order to explain the tension (see, for instance, \cite{2021A&ARv..29....9S} for more details). 

An interesting method to constrain the Hubble
constant is that  by using
galaxy cluster angular diameter distances obtained from the Sunyaev-Zel’dovich effect (SZE) plus X-ray observations \cite{2002ApJ...581...53R,2006ApJ...647...25B}. This is  independent of any calibrator usually adopted in the determinations of the
distance scale.
It is possible to take advantage of the different
electron density dependencies in these phenomena and evaluate the angular diameter distances of a galaxy cluster sample \cite{1999PhR...310...97B}. Along these lines,  Ref.\cite{2007MNRAS.379L...1C} combined the galaxy cluster SZE/X-ray data with  measurements of the baryon acoustic oscillation (BAO) and obtained $H_0 = 73.8^{+4.2}_{−3.3}$ in a flat $\Lambda$CDM model (only statistical errors). The Ref.\cite{2012JCAP...02..035H} considered a joint analysis involving the galaxy cluster SZE/X-ray data,  baryon acoustic oscillations (BAO) and the CMB Shift Parameter signature to obtain $H_0$ in more general dark energy models. For non-flat $\Lambda$CDM cosmologies $H_0 = 73.2^{+4.3}_{−3.7}$ km/s/Mpc, whereas for a flat universe with a constant dark energy equation of state parameter it was found $H_0 = 71.4^{+4.4}_{−3.4}$km/s/Mpc (only statistical errors)~\cite{2012JCAP...02..035H}. Thereafter, Ref.\cite{2014MNRAS.443L..74H} explored the robustness of the SZE/X-ray technique by  searching for systematic errors and its dependence from the cosmological model used. It was found that the $H_0$ value is very weakly dependent on the underlying cosmological model, but the morphology adopted to infer the distance to galaxy clusters changes the result considerably (see also \cite{2018BrJPh..48..521D}). Galaxy cluster X-ray gas mass fraction and the baryon acoustic oscillation measurements were also combined in order to obtain tight limits on $H_0$ considering the flat $\Lambda$CDM and XCDM models, and the non-flat $\Lambda$CDM model \cite{2020JCAP...09..053H}, the $H_0$ values obtained were:$H_0=65.9^{+1.5}_{-1.5}$ km/s/Mpc, $H_0=65.9^{+4.4}_{-4.0}$ km/s/Mpc and $H_0=64.3^{+ 4.5}_{- 4.4}$ km/s/Mpc in $2\sigma$ c.l., respectively, in full agreement with the Planck satellite results.

Particularly, in a  recent $H_0$ estimate from galaxy cluster observations, the authors of the Ref.\cite{2019A&A...621A..34K} evaluated this parameter by considering the following galaxy cluster scaling-relation, $Y_{SZE}D_{A}^{2}/C_{XSZ}Y_X = C $, where $Y_{SZE}D_{A}^{2}$ is the integrated Comptonization parameter of a galaxy cluster obtained via Sunyaev-Zeldovich (SZ) observations multiplied by its angular diameter distance,  $Y_X$ is the  X-ray counterpart, $C_{XSZ}$ is a constant and $C$ an arbitrary constant (if galaxy clusters are isothermal, $C$ would be exactly equal to unity, or constant with redshift if the galaxy clusters have a universal temperature profile) \cite{2006ApJ...650..128K,2010ApJ...715.1508S,2011MNRAS.416..801F,2012MNRAS.422.1999K,2012A&A...539A.120B}. The scaling-relations in galaxy clusters rise from the simplest model for the formation of structures, when gravity is the dominant process. In this scenario,  simple scaling-relations between basic galaxy cluster properties and the total mass are predicted by self-similar models\cite{1986MNRAS.222..323K}. By using 61 galaxy clusters with redshifts up to $z < 0.5$ \cite{2011A&A...536A..11P} observed with Planck and XMM-Newton, the authors from Ref.\cite{2019A&A...621A..34K} found $H_0 = 67\pm 3$ km/s/Mpc. However, it is  important to stress that the constant $C$ was obtained  from hydrodynamical simulations of massive clusters in a specific cosmological model (a flat $\Lambda$CDM model with $H_0=72$ km/s/Mpc and $\Omega_M=0.24$).

In this paper, we propose a new method to obtain $H_0$ by combining the galaxy cluster scaling-relation measurements, namely, $Y_{SZE}D_{A}^{2}/C_{XSZ}Y_X = C$ obtained from joint SZ/X-ray observations, with type Ia supernovae observations. Our method is divided into two parts: first, we consider  the cosmic distance relation validity $(D_A=D_L(1+z)^{-2})$,  SNe Ia observations and a sample of galaxy cluster scaling-relation measurements in low redshifts ($z<0.1$) to put observational limits on the $C$ ratio. {As we shall see  $C \equiv C(M_B)$. Therefore, we then consider a flat $\Lambda$CDM model for $D_{A}$, a Planck prior on $\Omega_M$ (the matter total density parameter),   a Gaussian prior on $M_B$ ($M_B = −19.253 \pm 0.027$) and put limits on $H_0$ by using the remaining galaxy cluster-scaling relation measurements with $z>0.1$ from the original sample. The complete galaxy cluster sample is  composed  of  61  $Y_{SZE} − Y_X$ measurements obtained from the first Planck mission all-sky data set \cite{2011A&A...536A..11P} jointly with deep XMM-Newton archive observations within the following redshift interval: $0.044 \leq z \leq  0.444$.  The fundamental idea of our method  is based on  the $C$ quantity to be constant with galaxy cluster redshift (or at least in the redshift range of the sample considered). As we shall see,  {we obtain:}  $ { H_0=73.014^{+7.435}_{-6.688}}$ km/s/Mpc, in full agreement with the latest results from {HST + SH0ES}. This result supports an observational verification of such scaling-relation from galaxy clusters.}

The manuscript is organized as follows. In Section~\ref{data}, we briefly explain the cosmological data sample used in our analysis. The methodology adopted in this work is presented in Section~\ref{methodology}. Section~\ref{sec:analysis} describes our analysis and results. We conclude in Section~\ref{sec:conclusions}.

\section{Cosmological data}
\label{data}
 
\begin{figure}
    \centering
    \includegraphics[width=10cm, height=6cm]{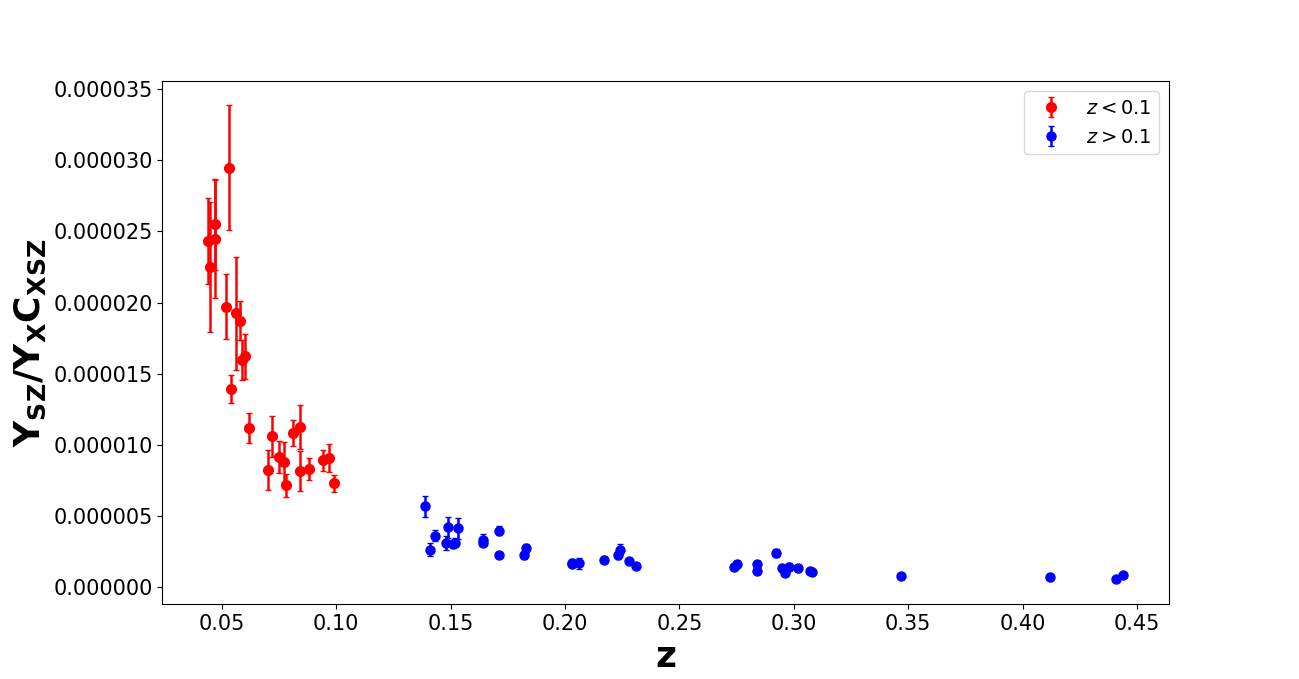} 
   \caption{61 $Y_{SZ}/Y_X C_{XSZ}$ galaxy clusters sample as a function of redshift\cite{2011A&A...536A..11P}. The $z<0.1$ sample are shown by red data points used to estimate the universal constant $C$. The blue points show the sample $z>0.1$ used to obtain the Hubble constant, $H_0$.}
    
    \label{fig:ratio}
\end{figure}

\begin{figure}
    \centering
    \includegraphics[width=10cm, height=6cm]{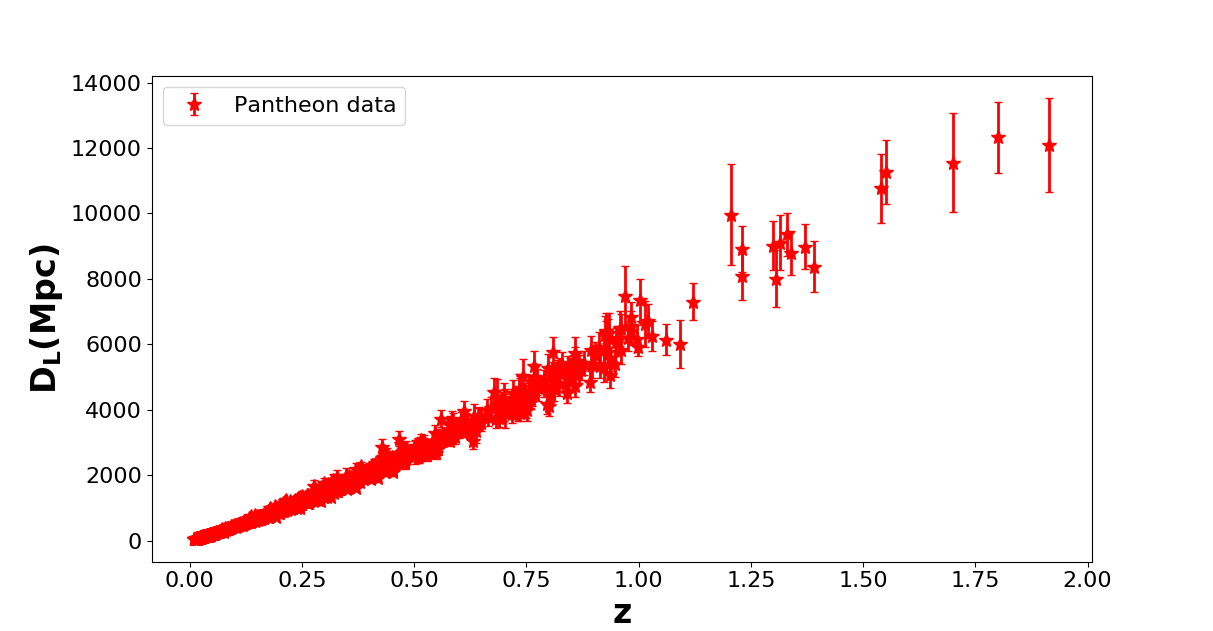} 
   \caption{The luminosity distance $D_L$, of full Pantheon(Type Ia SNe) sample as a function of redshift. {  For this plot we have assumed $M_B = −19.253 \pm 0.027$~\cite{riess22}.} }
    
    \label{fig:Pantheon}
\end{figure}

\begin{figure}
    \centering
    \includegraphics[width=10cm, height=6cm]{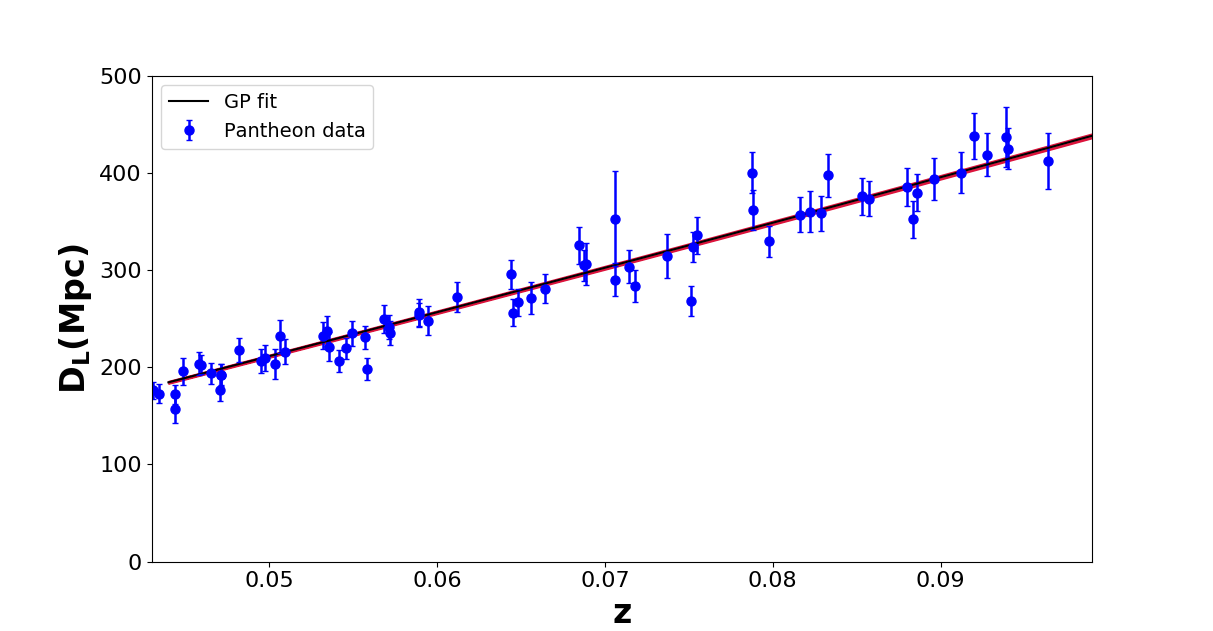} 
   \caption{The non-parametric reconstruction of luminosity distance as a function of $z$ at cluster's redshift. For this purpose, we use the Gaussian Processes Regression method~\cite{Haveesh}. The  black line along with the crimson shaded region shows the GP fit and $1\sigma$ error. The blue data points are Pantheon data in $z<0.1$. {  For this plot we have assumed $M_B = −19.253 \pm 0.027$~\cite{riess22}.}}
    
    \label{fig:DL}
\end{figure}

Our $H_0$ estimate is performed by using the following data set:

\begin{itemize}
\item  $Y_{SZE}-{Y_X}$  measurements of 61 galaxy clusters obtained from the first {\it Planck mission} all-sky data set jointly with deep XMM-Newton archive observations \cite{2011A&A...536A..11P} (see Fig~\ref{fig:ratio}). This is also known as the Planck-ESZ catalog.  This sample was detected at high  signal-to-noise  within the following redshift interval and mass, respectively: $0.044 \leq z \leq 0.444$ and $2 \mbox{x} 10^{14} M_{\odot} \leq M_{500} \leq 2 \mbox{x} 10^{15} M_{\odot}$, where $M_{500}$ is the total mass corresponding to a total density contrast of $500\rho_c(z)$, being $\rho_c(z)$  the critical density of the Universe at the cluster redshift. As it is largely known, one needs to add some complementary assumptions about the galaxy cluster physical properties in order to estimate their $Y_{SZE}-{Y_X}$  measurements. The thermal pressure ($P$) of the intra-cluster medium for each galaxy cluster used here was described by the Ref.~\cite{2011A&A...536A..11P} via the universal pressure profile discussed in details by the Ref.\cite{2010A&A...517A..92A}. This universal profile was obtained by comparing  the observational data  with simulated data, being the observational data  representative sample of nearby clusters covering the mass range $10^{14}M_{\odot} < M_{500} < 10^{15} M_{\odot}$). The $T_X$ quantity was measured in the $[0.15-0.75] R_{500}$ region. The Ref.~\cite{2011A&A...536A..11P} showed that the $\frac{Y_{SZE}D_{A}^{2}}{C_{XSZ}Y_X}$ ratio for  galaxy clusters considered in this work has very small scatter, at the level of $\approx 15\%$ (by using $D_A$ calculated from the {\it Planck mission} flat $\Lambda$CDM framework,). Moreover, it was also verified that this scaling-relation does not seem to depend crucially on the dynamical state of the clusters. Note that we used the same redshifts as provided in Planck ESZ papers~\cite{2011A&A...536A..11P}. Note however that some of these redshifts have been updated~\cite{Boraepj}.



\item SNe Ia: The Pantheon sample \cite{pantheon} considered in this work, is the most recent sample of SNe Ia consisting of 1049 spectroscopically confirmed data points and covering a redshift range of $0.01 \leq z \leq 2.3$ (see Fig~\ref{fig:Pantheon}). The luminosity distance, $D_L$ from its apparent magnitude ($ {m_B}$) and the absolute magnitude $M_B$ is given by the following relation

\begin{equation}
\label{MB_eq}
D_L=10^{( {m_B-M_B}-25)/5} \text{Mpc}.
\end{equation}
 {  Here we consider the absolute magnitude $M_B$ as a nuisance parameter and marginalize over $M_B$, while maximizing the Eq.~\ref{eq:logL2}. We use a Gaussian prior on $M_B = −19.253 \pm 0.027$, using  the latest local $H_0= 73.04 \pm 1.04$ km/sec/Mpc measurement from the HST + SH0ES team~\cite{riess22}
 (see also~\cite{george21,nunes21,Camarena_2021}).
 Moreover, in order to reconstruct the luminosity distance $D_L$ at each cluster's redshift, we use the Gaussian Process Regression (GPR). For this purpose, we used the {\tt scikit-learn} module in python~\cite{sklearn}. Gaussian Processes (GPs) offer a non-parametric way to model a function and are characterized by the mean function
and the kernel function~\cite{seikel12,benisty22}. For this work, we select the squared exponential covariance function, which is given by:
\begin{equation}
 \label{eq:kernal}
  K(x,\Tilde{x}) = \sigma_f^2 \exp  {\left[\frac{-(x-\Tilde{x})^2}{2l^2}\right]}, 
\end{equation}
It depends on two hyperparameters $\sigma_f$ and $l$ respectively where the length parameter $l$ controls the smoothness of the covariance function. Fig~\ref{fig:DL} shows the reconstructed luminosity distance as a function of $z$ using GPR.}

\end{itemize}
\vspace{0.2cm}

\section{Methodology}
\label{methodology}
\subsection{Galaxy cluster scaling-relation}
The Inverse Compton scattering between the CMB photons and the electrons present in the intra-cluster gas is characterized by a parameter called the integrated Compto-ionization parameter, $Y_{SZ}$~\cite{1999PhR...310...97B}.
$Y_{SZ}$ and its X-ray counterpart i.e. $Y_X$ both approximate the thermal energy of the intra-cluster gas within a cluster~\cite{More}.
It is well known that the ratio of $Y_{SZ}$ to $Y_X$ is nearly  constant as a function of redshift, as both of them scale with mass and redshift exactly in a  similar  manner ~\cite{2011A&A...536A..11P}. Therefore, both these ratios are different proxies for the thermal energy of the cluster~\cite{More}.

\begin{equation}
\frac{Y_{SZ} D_A^2 }{C_{XSZ} Y_X} = C
\label{eq:firstratio}
\end{equation}
where $C_{XSZ} \approx 1.416 \times 10^{-19} \left( \frac{ Mpc^2}{M_\odot keV}\right)$. $D_A$ is the angular diameter distance to the cluster. 
$C$ is an arbitrary constant which contains all the cluster's astrophysics. If galaxy clusters are isothermal then $C = 1$, and if galaxy clusters can be represented by an universal temperature profile then $C$ is expected to be a constant w.r.t redshift~\cite{loken02,galli,colaco19}. The simulations indicate a low scatter of $5-15\%$~\cite{planelles17,biffi14,fabjan11,kay12,stanek10}.
 
The quantity $Y_X$ depends on the intra-cluster gas mass, $M_g$ and also $Y_X \propto M_g \propto D_A^{3/2} D_L$. Again, if we consider a valid CDDR relation in a reference cosmology with $\Omega_m = 0.3$ and $H_0 = 70$ km/sec/Mpc then we get, $Y_X \propto (D_A^{ref})^{5/2}$. So in order to make our analysis independent of the reference cosmology, we multiply the $Y_X$ with $D_A^{5/2}$/ $(D_A^{ref})^{5/2}$. So from Eq~\ref{eq:firstratio}, we can write,

\begin{equation}
\frac{Y_{SZ} (D_A^{ref})^{5/2} }{C_{XSZ} Y_X D_A^{1/2}} = C.
\label{eq:ratioeq}
\end{equation}
As it is largely known,  the hierarchical structure formation theory results in galaxy cluster scaling-relations, where gravity is the dominant process. In other words, self-similar models predict simple scaling relations between basic galaxy cluster properties such as the above scaling relation \cite{1986MNRAS.222..323K}. {  It should be stressed however these relations are valid only if the condition of hydrostatic equilibrium
holds. This assumption also breaks down in disturbed systems undergoing mergers and neglects the effects of physical processes internal to the cluster such as feedback from active galactic nuclei and star formation.

In this line, a recent work \cite{2020JCAP...12..027H},  by using galaxy cluster observations, the cosmic distance duality relation validity and SNe Ia, through a Bayesian analysis  showed that other $C(z)$ functions (besides $C$ as an universal constant)  cannot be still discarded. Therefore,  an observational verification of the scaling-relations are still welcome  for galaxy cluster  cosmology. }

\begin{figure*}
    \centering
    \includegraphics[width=14cm, height=8cm]{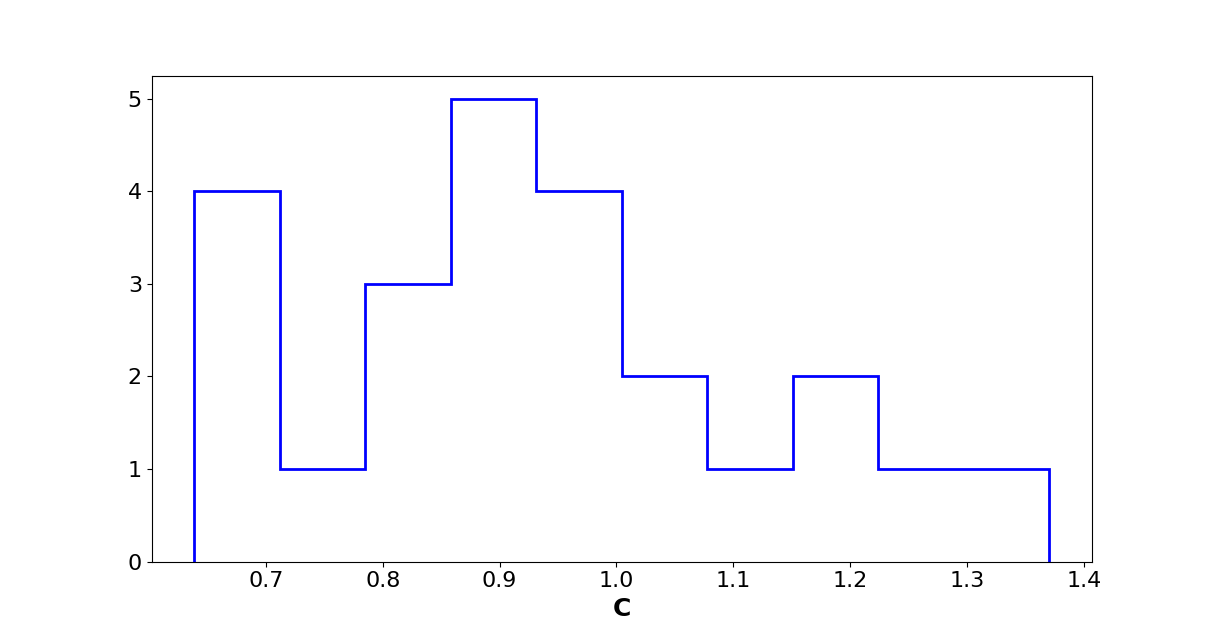} 
   \caption{The distribution of C with mean value of $ {C=0.93\pm0.16}$. {  For this plot we have assumed $M_B = −19.253 \pm 0.027$~\cite{riess22}.}}
    
    \label{fig:hist}
\end{figure*}

\subsection{Obtaining $C$ from a joint analysis with GC and SNe Ia}

If we consider the validity of duality relation(CDDR) then $D_L = (1+z)^2 D_A$~\cite{mendonca21_cddr} and the  SNe Ia observations, then Eq~\ref{eq:ratioeq} can be recast as,

\begin{equation}
C = \frac{Y_{SZ}(1+z)(D_A^{ref})^{5/2} }{C_{XSZ} Y_X D_{L,SNe}^{1/2}} .
\label{eq:ratio}
\end{equation}
 where $D_L$ is a function of $M_B$ as given by eq~\ref{MB_eq}. Therefore 
 \begin{equation}
     C \equiv C(M_B)
 \end{equation}
 
 Then, we use the galaxy cluster data with $z<0.1$ in order to obtain $C$. It is worth pointing out that the fundamental idea of our method  is based on  this quantity to be constant with galaxy cluster redshifts (or at least in the redshift range of the sample considered). Following the simulations, we have added a scatter of 15$\%$ in our analyses \cite{planelles17,biffi14,fabjan11,kay12,stanek10}. As commented before,  $D_L$ for each galaxy cluster  is estimated from the SNe Ia Pantheon sample via Eq~\ref{MB_eq} by applying the Gaussian Process Regression (see ~\cite{Haveesh} and references therein) at the cluster's redshifts (See Fig~\ref{fig:DL}). In Fig~\ref{fig:hist}, we plot the distribution of $C$ with mean value of $ {C=0.93\pm0.16}${(by assuming $ {M_B = −19.253 \pm0.027}$)}. Then, our result on $C$ indicates a departure from an isothermal assumption for the temperature profile of the galaxy clusters used in our analysis. 

\subsection{Obtaining $H_0$ from galaxy cluster scaling-relation}

In order to obtain the Hubble constant, we can return to Eq~\ref{eq:ratio} and consider a flat $\Lambda$CDM model to $D_L$, then,

\begin{equation}
(D^{\Lambda CDM}_{L})^{1/2} = \frac{Y_{SZ}(1+z)(D_A^{ref})^{5/2} }{C_{XSZ}Y_X C(M_B)}.
\end{equation}
Finally, substituting the observational $C$ value and considering 
\begin{equation}
D^{\Lambda CDM}_{L} (z) = \frac{c (1+z)}{H_0}\int_{0}^{z}\frac{dz^{'}}{\sqrt{(\Omega_m(1+z)^3 + (1-\Omega_m))}},
\label{eq:dl}
\end{equation}
one may estimate the $H_0$ value by using the galaxy cluster data from \cite{2011A&A...536A..11P}  with $z>0.1$.

\section{Analysis and Results}
\label{sec:analysis}

\begin{figure*}
    \centering
    \includegraphics[width=17cm, height=12cm]{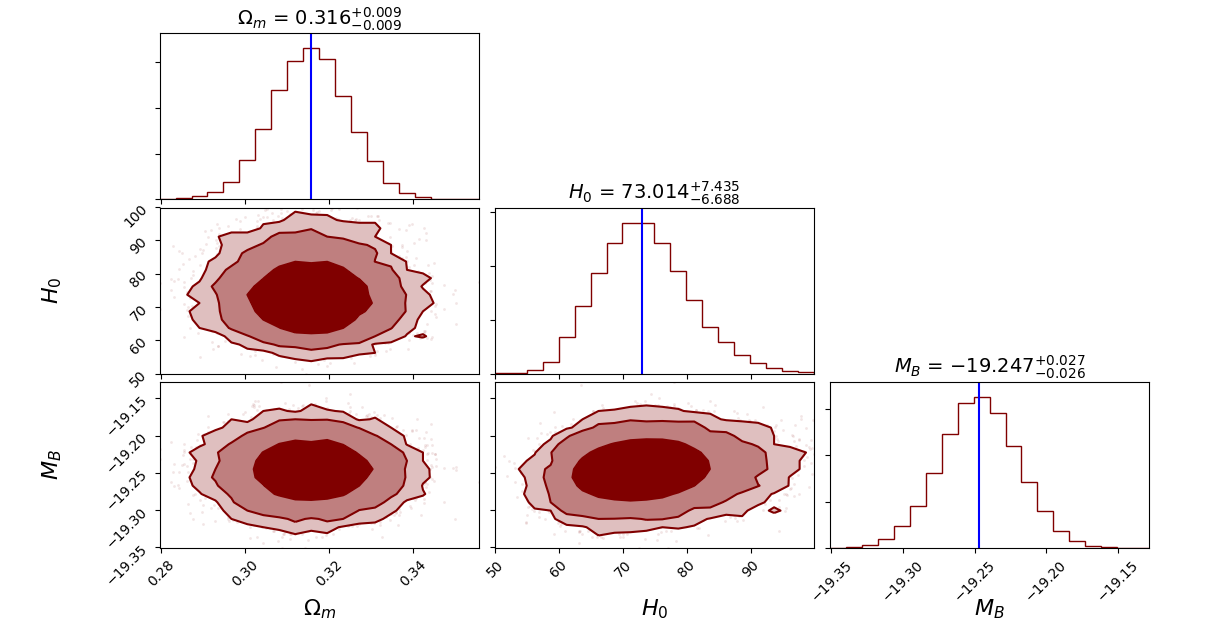} 
   \caption{The 1-D marginalized likelihood distributions along with 2-D marginalized constraints showing the 68\%, 95\%, and 99\% confidence regions for the parameters $\Omega_m$, $H_0$,{  and $M_B$,} obtained using the {\tt Corner} python module~\cite{corner}.}
    \label{fig:cornerplot}
\end{figure*}

The log-likelihood function used to estimate the parameters ($\Omega_m,H_0, {M_B}$) is given by,

\begin{widetext}
\begin{equation}
    \label{eq:logL2}
   -2\ln\mathcal{L} =  \sum_{i=1}^{n}\left[\frac{D_{L}^{\Lambda CDM}(\Omega_m,H_0, {M_B})^{1/2} - \frac{Y_{SZ}(1+z)(D_A^{ref})^{5/2} }{C_{XSZ} Y_X C(M_B)}}{\sigma_{i}} \right]^2 + \sum_{i=1}^{n} \ln 2\pi{\sigma_{i}^2}. 
\end{equation}  
\end{widetext} 
Here $\sigma_{i}^2$ stands for the observational errors in $Y_{SZ}, Y_X, D_A^{ref}$, $D_L$, $ {M_B}$ and $C$ which is calculated by the error propagation method. We use the $\tt{emcee}$ MCMC sampler~\cite{emcee} to maximize the log-likelihood(Eq~\ref{eq:logL2}). Here, we adopt a Gaussian prior for $\Omega_m$ ($\mu, \sigma$ = 0.3156, 0.0091) {  and $M_B$ ($\mu, \sigma$ = -19.253, 0.027) respectively whereas} a Uniform prior for $H_0$($40 \leq H_0 \leq 100$) {  is used}. Our main results are shown in Fig~\ref{fig:cornerplot}, which show the 68\%, 95\%, and 99\% confidence level along with the marginalized one-dimensional likelihoods for each of the parameters $(\Omega_m,H_0, {M_B})$. We report: $ {H_0 = 73.014^{+7.435}_{-6.688}}$ for the Planck ESZ sample. As one can see,  the $H_0$ values estimated from our analysis is consistent with the recent {  HST + SH0ES} $H_0$ value within $1\sigma$. Moreover, our result  points for the robustness of the SZE/X-ray measurements of galaxy clusters, as well as for an observational verification of the scaling-relation as given by the Eq~\ref{eq:firstratio}.


{
 \section{Comparing results}

 In this point, it is worth to compare our results with that one from the Ref.\cite{2019A&A...621A..34K}  where was obtained $H_0 = 67\pm 3$ km/s/Mpc using the same galaxy cluster data. As commented earlier, these authors considered  a $C$ value  obtained  from hydrodynamical simulations  in a specific cosmological model (a flat $\Lambda$CDM model with $H_0=72$ km/s/Mpc, $\Omega_M=0.24$ and a fraction of hydrogen mass $X = 0.76$). In their method, the simulations were necessary to taking into account possible biases induced, for example, by cluster triaxiality (it was indicate in the Ref.\cite{2011A&A...536A..11P} that possible effects of clumping in the X-ray gas were not significant for the present galaxy cluster data). However,  it is very important to stress that  the physical processes used in hydrodynamic simulations could do not span the entire range of physical processes allowed by our current understanding of the intra-cluster medium. Clearly,  our method can remove this limitation since only observational data are directly  used to obtain the $C$ parameter. On the other hand, there is much observational evidence for galaxy clusters not be spherical objects,  which is a worth current limiting factor of our method.  Different authors have proposed to combine complementary data sets to reconstruct the three dimensional properties of galaxy clusters (gravitational lensing, X-ray and SZE observations) (see details in the Ref.\cite{2013SSRv..177..155L}). Progress in this direction could lead our method to obtain a more accurate value for the Hubble constant.  Then, with the systematic errors under control, discrepant results from our method and the hydrodynamical
simulations may indicate the presence of some unknown physical
mechanism in the intra-cluster medium not yet considered
in the simulations.
}

\section{Conclusions}
\label{sec:conclusions}
We proposed a  method for determining $H_0$ using joint SZ/X-Ray observations of galaxy clusters in conjunction with Type Ia SNe. By using galaxy cluster data with $z<0.1$, we first determine the ratio of the integrated SZ compton-ionization parameter to its X-ray counterpart (cf. Eq~\ref{eq:firstratio}) considering the cosmic distance duality relation to express the angular diameter distance, which is present in the aforementioned ratio to the luminosity distance ($D_L$) (see Eq~\ref{MB_eq}). At each cluster redshift, we obtained an independent estimate of $D_L$ using Type Ia supernova and an interpolation technique (see Fig~\ref{fig:DL}). { Here, we have used the absolute magnitude $M_B$ as a nuisance parameter and adopted a Gaussian prior($\mu, \sigma$ = -19.253, 0.027) while maximizing the likelihood.} From this step we obtained: $ {C=0.93\pm0.16}$ (see Fig~\ref{fig:hist}). For this purpose, we used a subsample from a original galaxy cluster  data consisting of 61 Planck ESZ clusters in the redshift range $0.044 \leq z \leq 0.444$ (see Fig~\ref{fig:ratio}). 

Then, we considered a flat $\Lambda$CDM model for $D_{A}$, a Planck prior on $\Omega_M$ (the matter total density parameter) and put limits on $H_0$ by using the remaining galaxy cluster-scaling relation measurements with $z>0.1$ from the original sample and the $C$ value obtained from first step (where we used galaxy clusters with $z<0.1$). As one may see, the key assumption of our method  is based on  the $C$ quantity to be constant in the redshift range of the sample considered. Finally, by using the galaxy cluster data in higher redshifts ($z>0.1$), {  we obtained} $ { H_0=73.014^{+7.435}_{-6.688}}$ km/s/Mpc (see Fig~\ref{fig:cornerplot}), which represents 9.7\% accuracy measurement on the Hubble constant.

  As it is largely known, it
is needed to deeply know some intra-cluster gas and dark
matter properties in order to use galaxy clusters as a cosmological probe. Key ingredients for analyses that aim to constrain cosmological parameters are the scaling-relations between the observable properties and the total masses of these structures, so they need to be well-calibrated. Therefore,  our self-consistent test  points for the robustness of the SZE/X-ray measurements of galaxy clusters used in analyses, as well as for an observational verification of the scaling-relation as given by the Eq~\ref{eq:firstratio}.
Our method did not depend on any hydrodynamic simulations in a specific cosmological model to obtain the $C$ and $H_0$ parameters. 

{ It is worth to comment that the future eROSITA observations will provide significant gains over available X-ray surveys, where $\approx$ 100,000 galaxy clusters are expected to be detected in X-ray band \cite{2017A&A...606A.118H}. On the other hand, surveys on mm band  have been performed is last years (ACT, SPT and Planck mission),  with the promise of  more to come (see, for instance, NIKA2 ESZ Large Program\footnote{The NIKA2 ESZ Large is a Program dedicated to the  ESZ mapping of galaxy clusters in high redshift. It is  expected to deliver high  quality of $Y - M$ scaling relation measurements.} \cite{2022EPJWC.25700038P,2022EPJWC.25700025K}). { Finally, as one may see, the spherical hypothesis describing the morphology of galaxy clusters is a limiting factor of our method. Through the years,  different studies have shown that by combining complementary data sets it is possible to reconstruct the three dimensional properties of galaxy clusters (gravitational lensing, X-ray and SZE observations). Then, progress in this direction could lead our method to obtain a more accurate value for the Hubble constant and also indicate the presence of some unknown physical mechanism in the intra-cluster medium not yet considered in the simulations. Therefore, we hope that the method proposed here can be performed with higher quality data in the near future.}

  .

}

\section*{ACKNOWLEDGEMENT}
We would like to thank Dr. Shantanu Desai for his valuable comments on this manuscript. KB acknowledges the Department of Science and Technology, Government of India for providing the financial support under DST-INSPIRE Fellowship program. RFLH
thanks CNPq No.428755/2018-6 and 305930/2017-6. {We are grateful to the anonymous referees for useful constructive feedback
on our manuscript.}

\bibliography{ref}

\begin{thebibliography}{61}
\expandafter\ifx\csname natexlab\endcsname\relax\def\natexlab#1{#1}\fi
\expandafter\ifx\csname bibnamefont\endcsname\relax
  \def\bibnamefont#1{#1}\fi
\expandafter\ifx\csname bibfnamefont\endcsname\relax
  \def\bibfnamefont#1{#1}\fi
\expandafter\ifx\csname citenamefont\endcsname\relax
  \def\citenamefont#1{#1}\fi
\expandafter\ifx\csname url\endcsname\relax
  \def\url#1{\texttt{#1}}\fi
\expandafter\ifx\csname urlprefix\endcsname\relax\def\urlprefix{URL }\fi
\providecommand{\bibinfo}[2]{#2}
\providecommand{\eprint}[2][]{\url{#2}}

\bibitem[{\citenamefont{{Weinberg} et~al.}(2013)\citenamefont{{Weinberg},
  {Mortonson}, {Eisenstein}, {Hirata}, {Riess}, and {Rozo}}}]{Weinberg2013}
\bibinfo{author}{\bibfnamefont{D.~H.} \bibnamefont{{Weinberg}}},
  \bibinfo{author}{\bibfnamefont{M.~J.} \bibnamefont{{Mortonson}}},
  \bibinfo{author}{\bibfnamefont{D.~J.} \bibnamefont{{Eisenstein}}},
  \bibinfo{author}{\bibfnamefont{C.}~\bibnamefont{{Hirata}}},
  \bibinfo{author}{\bibfnamefont{A.~G.} \bibnamefont{{Riess}}},
  \bibnamefont{and} \bibinfo{author}{\bibfnamefont{E.}~\bibnamefont{{Rozo}}},
  \bibinfo{journal}{\physrep} \textbf{\bibinfo{volume}{530}},
  \bibinfo{pages}{87} (\bibinfo{year}{2013}), \eprint{1201.2434}.

\bibitem[{\citenamefont{{Caldwell} and {Kamionkowski}}(2009)}]{Caldwell2009}
\bibinfo{author}{\bibfnamefont{R.~R.} \bibnamefont{{Caldwell}}}
  \bibnamefont{and}
  \bibinfo{author}{\bibfnamefont{M.}~\bibnamefont{{Kamionkowski}}},
  \bibinfo{journal}{Annual Review of Nuclear and Particle Science}
  \textbf{\bibinfo{volume}{59}}, \bibinfo{pages}{397} (\bibinfo{year}{2009}),
  \eprint{0903.0866}.

\bibitem[{\citenamefont{{Huterer} and {Shafer}}(2018)}]{Huterer16}
\bibinfo{author}{\bibfnamefont{D.}~\bibnamefont{{Huterer}}} \bibnamefont{and}
  \bibinfo{author}{\bibfnamefont{D.~L.} \bibnamefont{{Shafer}}},
  \bibinfo{journal}{Reports on Progress in Physics}
  \textbf{\bibinfo{volume}{81}}, \bibinfo{eid}{016901} (\bibinfo{year}{2018}),
  \eprint{1709.01091}.

\bibitem[{\citenamefont{{Suyu} et~al.}(2012)\citenamefont{{Suyu}, {Treu},
  {Blandford}, {Freedman}, {Hilbert}, {Blake}, {Braatz}, {Courbin}, {Dunkley},
  {Greenhill} et~al.}}]{2012arXiv1202.4459S}
\bibinfo{author}{\bibfnamefont{S.~H.} \bibnamefont{{Suyu}}},
  \bibinfo{author}{\bibfnamefont{T.}~\bibnamefont{{Treu}}},
  \bibinfo{author}{\bibfnamefont{R.~D.} \bibnamefont{{Blandford}}},
  \bibinfo{author}{\bibfnamefont{W.~L.} \bibnamefont{{Freedman}}},
  \bibinfo{author}{\bibfnamefont{S.}~\bibnamefont{{Hilbert}}},
  \bibinfo{author}{\bibfnamefont{C.}~\bibnamefont{{Blake}}},
  \bibinfo{author}{\bibfnamefont{J.}~\bibnamefont{{Braatz}}},
  \bibinfo{author}{\bibfnamefont{F.}~\bibnamefont{{Courbin}}},
  \bibinfo{author}{\bibfnamefont{J.}~\bibnamefont{{Dunkley}}},
  \bibinfo{author}{\bibfnamefont{L.}~\bibnamefont{{Greenhill}}},
  \bibnamefont{et~al.}, \bibinfo{journal}{arXiv e-prints}
  \bibinfo{eid}{arXiv:1202.4459} (\bibinfo{year}{2012}), \eprint{1202.4459}.

\bibitem[{\citenamefont{{Shah} et~al.}(2021)\citenamefont{{Shah}, {Lemos}, and
  {Lahav}}}]{2021A&ARv..29....9S}
\bibinfo{author}{\bibfnamefont{P.}~\bibnamefont{{Shah}}},
  \bibinfo{author}{\bibfnamefont{P.}~\bibnamefont{{Lemos}}}, \bibnamefont{and}
  \bibinfo{author}{\bibfnamefont{O.}~\bibnamefont{{Lahav}}},
  \bibinfo{journal}{\aapr} \textbf{\bibinfo{volume}{29}}, \bibinfo{eid}{9}
  (\bibinfo{year}{2021}), \eprint{2109.01161}.

\bibitem[{\citenamefont{{Planck Collaboration}
  et~al.}(2020)\citenamefont{{Planck Collaboration}, {Aghanim}, {Akrami},
  {Ashdown}, {Aumont}, {Baccigalupi}, {Ballardini}, {Banday}, {Barreiro},
  {Bartolo} et~al.}}]{2020A&A...641A...6P}
\bibinfo{author}{\bibnamefont{{Planck Collaboration}}},
  \bibinfo{author}{\bibfnamefont{N.}~\bibnamefont{{Aghanim}}},
  \bibinfo{author}{\bibfnamefont{Y.}~\bibnamefont{{Akrami}}},
  \bibinfo{author}{\bibfnamefont{M.}~\bibnamefont{{Ashdown}}},
  \bibinfo{author}{\bibfnamefont{J.}~\bibnamefont{{Aumont}}},
  \bibinfo{author}{\bibfnamefont{C.}~\bibnamefont{{Baccigalupi}}},
  \bibinfo{author}{\bibfnamefont{M.}~\bibnamefont{{Ballardini}}},
  \bibinfo{author}{\bibfnamefont{A.~J.} \bibnamefont{{Banday}}},
  \bibinfo{author}{\bibfnamefont{R.~B.} \bibnamefont{{Barreiro}}},
  \bibinfo{author}{\bibfnamefont{N.}~\bibnamefont{{Bartolo}}},
  \bibnamefont{et~al.}, \bibinfo{journal}{\aap} \textbf{\bibinfo{volume}{641}},
  \bibinfo{eid}{A6} (\bibinfo{year}{2020}), \eprint{1807.06209}.

\bibitem[{\citenamefont{{Riess} et~al.}(2021)\citenamefont{{Riess}, {Yuan},
  {Macri}, {Scolnic}, {Brout}, {Casertano}, {Jones}, {Murakami}, {Breuval},
  {Brink} et~al.}}]{2021arXiv211204510R}
\bibinfo{author}{\bibfnamefont{A.~G.} \bibnamefont{{Riess}}},
  \bibinfo{author}{\bibfnamefont{W.}~\bibnamefont{{Yuan}}},
  \bibinfo{author}{\bibfnamefont{L.~M.} \bibnamefont{{Macri}}},
  \bibinfo{author}{\bibfnamefont{D.}~\bibnamefont{{Scolnic}}},
  \bibinfo{author}{\bibfnamefont{D.}~\bibnamefont{{Brout}}},
  \bibinfo{author}{\bibfnamefont{S.}~\bibnamefont{{Casertano}}},
  \bibinfo{author}{\bibfnamefont{D.~O.} \bibnamefont{{Jones}}},
  \bibinfo{author}{\bibfnamefont{Y.}~\bibnamefont{{Murakami}}},
  \bibinfo{author}{\bibfnamefont{L.}~\bibnamefont{{Breuval}}},
  \bibinfo{author}{\bibfnamefont{T.~G.} \bibnamefont{{Brink}}},
  \bibnamefont{et~al.}, \bibinfo{journal}{arXiv e-prints}
  \bibinfo{eid}{arXiv:2112.04510} (\bibinfo{year}{2021}), \eprint{2112.04510}.

\bibitem[{\citenamefont{{Wong} et~al.}(2020)\citenamefont{{Wong}, {Suyu},
  {Chen}, {Rusu}, {Millon}, {Sluse}, {Bonvin}, {Fassnacht}, {Taubenberger},
  {Auger} et~al.}}]{2020MNRAS.498.1420W}
\bibinfo{author}{\bibfnamefont{K.~C.} \bibnamefont{{Wong}}},
  \bibinfo{author}{\bibfnamefont{S.~H.} \bibnamefont{{Suyu}}},
  \bibinfo{author}{\bibfnamefont{G.~C.~F.} \bibnamefont{{Chen}}},
  \bibinfo{author}{\bibfnamefont{C.~E.} \bibnamefont{{Rusu}}},
  \bibinfo{author}{\bibfnamefont{M.}~\bibnamefont{{Millon}}},
  \bibinfo{author}{\bibfnamefont{D.}~\bibnamefont{{Sluse}}},
  \bibinfo{author}{\bibfnamefont{V.}~\bibnamefont{{Bonvin}}},
  \bibinfo{author}{\bibfnamefont{C.~D.} \bibnamefont{{Fassnacht}}},
  \bibinfo{author}{\bibfnamefont{S.}~\bibnamefont{{Taubenberger}}},
  \bibinfo{author}{\bibfnamefont{M.~W.} \bibnamefont{{Auger}}},
  \bibnamefont{et~al.}, \bibinfo{journal}{\mnras}
  \textbf{\bibinfo{volume}{498}}, \bibinfo{pages}{1420} (\bibinfo{year}{2020}),
  \eprint{1907.04869}.

\bibitem[{\citenamefont{{Gariazzo} et~al.}(2021)\citenamefont{{Gariazzo}, {Di
  Valentino}, {Mena}, and {Nunes}}}]{2021arXiv211103152G}
\bibinfo{author}{\bibfnamefont{S.}~\bibnamefont{{Gariazzo}}},
  \bibinfo{author}{\bibfnamefont{E.}~\bibnamefont{{Di Valentino}}},
  \bibinfo{author}{\bibfnamefont{O.}~\bibnamefont{{Mena}}}, \bibnamefont{and}
  \bibinfo{author}{\bibfnamefont{R.~C.} \bibnamefont{{Nunes}}},
  \bibinfo{journal}{arXiv e-prints} \bibinfo{eid}{arXiv:2111.03152}
  (\bibinfo{year}{2021}), \eprint{2111.03152}.

\bibitem[{\citenamefont{{Bethapudi} and {Desai}}(2017)}]{Bethapudi}
\bibinfo{author}{\bibfnamefont{S.}~\bibnamefont{{Bethapudi}}} \bibnamefont{and}
  \bibinfo{author}{\bibfnamefont{S.}~\bibnamefont{{Desai}}},
  \bibinfo{journal}{European Physical Journal Plus}
  \textbf{\bibinfo{volume}{132}}, \bibinfo{eid}{78} (\bibinfo{year}{2017}),
  \eprint{1701.01789}.

\bibitem[{\citenamefont{{Di Valentino} et~al.}(2021)\citenamefont{{Di
  Valentino}, {Mena}, {Pan}, {Visinelli}, {Yang}, {Melchiorri}, {Mota},
  {Riess}, and {Silk}}}]{DiValentino21}
\bibinfo{author}{\bibfnamefont{E.}~\bibnamefont{{Di Valentino}}},
  \bibinfo{author}{\bibfnamefont{O.}~\bibnamefont{{Mena}}},
  \bibinfo{author}{\bibfnamefont{S.}~\bibnamefont{{Pan}}},
  \bibinfo{author}{\bibfnamefont{L.}~\bibnamefont{{Visinelli}}},
  \bibinfo{author}{\bibfnamefont{W.}~\bibnamefont{{Yang}}},
  \bibinfo{author}{\bibfnamefont{A.}~\bibnamefont{{Melchiorri}}},
  \bibinfo{author}{\bibfnamefont{D.~F.} \bibnamefont{{Mota}}},
  \bibinfo{author}{\bibfnamefont{A.~G.} \bibnamefont{{Riess}}},
  \bibnamefont{and} \bibinfo{author}{\bibfnamefont{J.}~\bibnamefont{{Silk}}},
  \bibinfo{journal}{Classical and Quantum Gravity}
  \textbf{\bibinfo{volume}{38}}, \bibinfo{eid}{153001} (\bibinfo{year}{2021}),
  \eprint{2103.01183}.

\bibitem[{\citenamefont{{Sch{\"o}neberg}
  et~al.}(2021)\citenamefont{{Sch{\"o}neberg}, {Abell{\'a}n}, {P{\'e}rez
  S{\'a}nchez}, {Witte}, {Poulin}, and {Lesgourgues}}}]{Julien}
\bibinfo{author}{\bibfnamefont{N.}~\bibnamefont{{Sch{\"o}neberg}}},
  \bibinfo{author}{\bibfnamefont{G.~F.} \bibnamefont{{Abell{\'a}n}}},
  \bibinfo{author}{\bibfnamefont{A.}~\bibnamefont{{P{\'e}rez S{\'a}nchez}}},
  \bibinfo{author}{\bibfnamefont{S.~J.} \bibnamefont{{Witte}}},
  \bibinfo{author}{\bibfnamefont{c.~V.} \bibnamefont{{Poulin}}},
  \bibnamefont{and}
  \bibinfo{author}{\bibfnamefont{J.}~\bibnamefont{{Lesgourgues}}},
  \bibinfo{journal}{arXiv e-prints} \bibinfo{eid}{arXiv:2107.10291}
  (\bibinfo{year}{2021}), \eprint{2107.10291}.

\bibitem[{\citenamefont{{Verde} et~al.}(2019)\citenamefont{{Verde}, {Treu}, and
  {Riess}}}]{Verde}
\bibinfo{author}{\bibfnamefont{L.}~\bibnamefont{{Verde}}},
  \bibinfo{author}{\bibfnamefont{T.}~\bibnamefont{{Treu}}}, \bibnamefont{and}
  \bibinfo{author}{\bibfnamefont{A.~G.} \bibnamefont{{Riess}}},
  \bibinfo{journal}{Nature Astronomy} \textbf{\bibinfo{volume}{3}},
  \bibinfo{pages}{891} (\bibinfo{year}{2019}), \eprint{1907.10625}.

\bibitem[{\citenamefont{{Freedman}}(2021)}]{Freedman21}
\bibinfo{author}{\bibfnamefont{W.~L.} \bibnamefont{{Freedman}}},
  \bibinfo{journal}{\apj} \textbf{\bibinfo{volume}{919}}, \bibinfo{eid}{16}
  (\bibinfo{year}{2021}), \eprint{2106.15656}.

\bibitem[{\citenamefont{{M{\"o}rtsell} and
  {Dhawan}}(2018)}]{2018JCAP...09..025M}
\bibinfo{author}{\bibfnamefont{E.}~\bibnamefont{{M{\"o}rtsell}}}
  \bibnamefont{and} \bibinfo{author}{\bibfnamefont{S.}~\bibnamefont{{Dhawan}}},
  \bibinfo{journal}{\jcap} \textbf{\bibinfo{volume}{2018}}, \bibinfo{eid}{025}
  (\bibinfo{year}{2018}), \eprint{1801.07260}.

\bibitem[{\citenamefont{{Bolejko}}(2018)}]{2018PhRvD..97j3529B}
\bibinfo{author}{\bibfnamefont{K.}~\bibnamefont{{Bolejko}}},
  \bibinfo{journal}{\prd} \textbf{\bibinfo{volume}{97}}, \bibinfo{eid}{103529}
  (\bibinfo{year}{2018}), \eprint{1712.02967}.

\bibitem[{\citenamefont{{D'Eramo} et~al.}(2018)\citenamefont{{D'Eramo},
  {Ferreira}, {Notari}, and {Bernal}}}]{2018JCAP...11..014D}
\bibinfo{author}{\bibfnamefont{F.}~\bibnamefont{{D'Eramo}}},
  \bibinfo{author}{\bibfnamefont{R.~Z.} \bibnamefont{{Ferreira}}},
  \bibinfo{author}{\bibfnamefont{A.}~\bibnamefont{{Notari}}}, \bibnamefont{and}
  \bibinfo{author}{\bibfnamefont{J.~L.} \bibnamefont{{Bernal}}},
  \bibinfo{journal}{\jcap} \textbf{\bibinfo{volume}{2018}}, \bibinfo{eid}{014}
  (\bibinfo{year}{2018}), \eprint{1808.07430}.

\bibitem[{\citenamefont{{Reese} et~al.}(2002)\citenamefont{{Reese},
  {Carlstrom}, {Joy}, {Mohr}, {Grego}, and {Holzapfel}}}]{2002ApJ...581...53R}
\bibinfo{author}{\bibfnamefont{E.~D.} \bibnamefont{{Reese}}},
  \bibinfo{author}{\bibfnamefont{J.~E.} \bibnamefont{{Carlstrom}}},
  \bibinfo{author}{\bibfnamefont{M.}~\bibnamefont{{Joy}}},
  \bibinfo{author}{\bibfnamefont{J.~J.} \bibnamefont{{Mohr}}},
  \bibinfo{author}{\bibfnamefont{L.}~\bibnamefont{{Grego}}}, \bibnamefont{and}
  \bibinfo{author}{\bibfnamefont{W.~L.} \bibnamefont{{Holzapfel}}},
  \bibinfo{journal}{\apj} \textbf{\bibinfo{volume}{581}}, \bibinfo{pages}{53}
  (\bibinfo{year}{2002}), \eprint{astro-ph/0205350}.

\bibitem[{\citenamefont{{Bonamente} et~al.}(2006)\citenamefont{{Bonamente},
  {Joy}, {LaRoque}, {Carlstrom}, {Reese}, and {Dawson}}}]{2006ApJ...647...25B}
\bibinfo{author}{\bibfnamefont{M.}~\bibnamefont{{Bonamente}}},
  \bibinfo{author}{\bibfnamefont{M.~K.} \bibnamefont{{Joy}}},
  \bibinfo{author}{\bibfnamefont{S.~J.} \bibnamefont{{LaRoque}}},
  \bibinfo{author}{\bibfnamefont{J.~E.} \bibnamefont{{Carlstrom}}},
  \bibinfo{author}{\bibfnamefont{E.~D.} \bibnamefont{{Reese}}},
  \bibnamefont{and} \bibinfo{author}{\bibfnamefont{K.~S.}
  \bibnamefont{{Dawson}}}, \bibinfo{journal}{\apj}
  \textbf{\bibinfo{volume}{647}}, \bibinfo{pages}{25} (\bibinfo{year}{2006}),
  \eprint{astro-ph/0512349}.

\bibitem[{\citenamefont{{Birkinshaw}}(1999)}]{1999PhR...310...97B}
\bibinfo{author}{\bibfnamefont{M.}~\bibnamefont{{Birkinshaw}}},
  \bibinfo{journal}{\physrep} \textbf{\bibinfo{volume}{310}},
  \bibinfo{pages}{97} (\bibinfo{year}{1999}), \eprint{astro-ph/9808050}.

\bibitem[{\citenamefont{{Cunha} et~al.}(2007)\citenamefont{{Cunha}, {Marassi},
  and {Lima}}}]{2007MNRAS.379L...1C}
\bibinfo{author}{\bibfnamefont{J.~V.} \bibnamefont{{Cunha}}},
  \bibinfo{author}{\bibfnamefont{L.}~\bibnamefont{{Marassi}}},
  \bibnamefont{and} \bibinfo{author}{\bibfnamefont{J.~A.~S.}
  \bibnamefont{{Lima}}}, \bibinfo{journal}{\mnras}
  \textbf{\bibinfo{volume}{379}}, \bibinfo{pages}{L1} (\bibinfo{year}{2007}),
  \eprint{astro-ph/0611934}.

\bibitem[{\citenamefont{{Holanda} et~al.}(2012)\citenamefont{{Holanda},
  {Cunha}, {Marassi}, and {Lima}}}]{2012JCAP...02..035H}
\bibinfo{author}{\bibfnamefont{R.~F.~L.} \bibnamefont{{Holanda}}},
  \bibinfo{author}{\bibfnamefont{J.~V.} \bibnamefont{{Cunha}}},
  \bibinfo{author}{\bibfnamefont{L.}~\bibnamefont{{Marassi}}},
  \bibnamefont{and} \bibinfo{author}{\bibfnamefont{J.~A.~S.}
  \bibnamefont{{Lima}}}, \bibinfo{journal}{\jcap}
  \textbf{\bibinfo{volume}{2012}}, \bibinfo{eid}{035} (\bibinfo{year}{2012}),
  \eprint{1006.4200}.

\bibitem[{\citenamefont{{Holanda} et~al.}(2014)\citenamefont{{Holanda},
  {Busti}, and {Pordeus da Silva}}}]{2014MNRAS.443L..74H}
\bibinfo{author}{\bibfnamefont{R.~F.~L.} \bibnamefont{{Holanda}}},
  \bibinfo{author}{\bibfnamefont{V.~C.} \bibnamefont{{Busti}}},
  \bibnamefont{and} \bibinfo{author}{\bibfnamefont{G.}~\bibnamefont{{Pordeus da
  Silva}}}, \bibinfo{journal}{\mnras} \textbf{\bibinfo{volume}{443}},
  \bibinfo{pages}{L74} (\bibinfo{year}{2014}), \eprint{1404.4418}.

\bibitem[{\citenamefont{{da Silva} and
  {Cavalcanti}}(2018)}]{2018BrJPh..48..521D}
\bibinfo{author}{\bibfnamefont{G.~P.} \bibnamefont{{da Silva}}}
  \bibnamefont{and} \bibinfo{author}{\bibfnamefont{A.~G.}
  \bibnamefont{{Cavalcanti}}}, \bibinfo{journal}{Brazilian Journal of Physics}
  \textbf{\bibinfo{volume}{48}}, \bibinfo{pages}{521} (\bibinfo{year}{2018}),
  \eprint{1805.06849}.

\bibitem[{\citenamefont{{Holanda} et~al.}(2020)\citenamefont{{Holanda},
  {Pordeus-da-Silva}, and {Pereira}}}]{2020JCAP...09..053H}
\bibinfo{author}{\bibfnamefont{R.~F.~L.} \bibnamefont{{Holanda}}},
  \bibinfo{author}{\bibfnamefont{G.}~\bibnamefont{{Pordeus-da-Silva}}},
  \bibnamefont{and} \bibinfo{author}{\bibfnamefont{S.~H.}
  \bibnamefont{{Pereira}}}, \bibinfo{journal}{\jcap}
  \textbf{\bibinfo{volume}{2020}}, \bibinfo{eid}{053} (\bibinfo{year}{2020}),
  \eprint{2006.06712}.

\bibitem[{\citenamefont{{Kozmanyan} et~al.}(2019)\citenamefont{{Kozmanyan},
  {Bourdin}, {Mazzotta}, {Rasia}, and {Sereno}}}]{2019A&A...621A..34K}
\bibinfo{author}{\bibfnamefont{A.}~\bibnamefont{{Kozmanyan}}},
  \bibinfo{author}{\bibfnamefont{H.}~\bibnamefont{{Bourdin}}},
  \bibinfo{author}{\bibfnamefont{P.}~\bibnamefont{{Mazzotta}}},
  \bibinfo{author}{\bibfnamefont{E.}~\bibnamefont{{Rasia}}}, \bibnamefont{and}
  \bibinfo{author}{\bibfnamefont{M.}~\bibnamefont{{Sereno}}},
  \bibinfo{journal}{\aap} \textbf{\bibinfo{volume}{621}}, \bibinfo{eid}{A34}
  (\bibinfo{year}{2019}), \eprint{1809.09560}.

\bibitem[{\citenamefont{{Kravtsov} et~al.}(2006)\citenamefont{{Kravtsov},
  {Vikhlinin}, and {Nagai}}}]{2006ApJ...650..128K}
\bibinfo{author}{\bibfnamefont{A.~V.} \bibnamefont{{Kravtsov}}},
  \bibinfo{author}{\bibfnamefont{A.}~\bibnamefont{{Vikhlinin}}},
  \bibnamefont{and} \bibinfo{author}{\bibfnamefont{D.}~\bibnamefont{{Nagai}}},
  \bibinfo{journal}{\apj} \textbf{\bibinfo{volume}{650}}, \bibinfo{pages}{128}
  (\bibinfo{year}{2006}), \eprint{astro-ph/0603205}.

\bibitem[{\citenamefont{{Stanek}
  et~al.}(2010{\natexlab{a}})\citenamefont{{Stanek}, {Rasia}, {Evrard},
  {Pearce}, and {Gazzola}}}]{2010ApJ...715.1508S}
\bibinfo{author}{\bibfnamefont{R.}~\bibnamefont{{Stanek}}},
  \bibinfo{author}{\bibfnamefont{E.}~\bibnamefont{{Rasia}}},
  \bibinfo{author}{\bibfnamefont{A.~E.} \bibnamefont{{Evrard}}},
  \bibinfo{author}{\bibfnamefont{F.}~\bibnamefont{{Pearce}}}, \bibnamefont{and}
  \bibinfo{author}{\bibfnamefont{L.}~\bibnamefont{{Gazzola}}},
  \bibinfo{journal}{\apj} \textbf{\bibinfo{volume}{715}}, \bibinfo{pages}{1508}
  (\bibinfo{year}{2010}{\natexlab{a}}), \eprint{0910.1599}.

\bibitem[{\citenamefont{{Fabjan}
  et~al.}(2011{\natexlab{a}})\citenamefont{{Fabjan}, {Borgani}, {Rasia},
  {Bonafede}, {Dolag}, {Murante}, and {Tornatore}}}]{2011MNRAS.416..801F}
\bibinfo{author}{\bibfnamefont{D.}~\bibnamefont{{Fabjan}}},
  \bibinfo{author}{\bibfnamefont{S.}~\bibnamefont{{Borgani}}},
  \bibinfo{author}{\bibfnamefont{E.}~\bibnamefont{{Rasia}}},
  \bibinfo{author}{\bibfnamefont{A.}~\bibnamefont{{Bonafede}}},
  \bibinfo{author}{\bibfnamefont{K.}~\bibnamefont{{Dolag}}},
  \bibinfo{author}{\bibfnamefont{G.}~\bibnamefont{{Murante}}},
  \bibnamefont{and}
  \bibinfo{author}{\bibfnamefont{L.}~\bibnamefont{{Tornatore}}},
  \bibinfo{journal}{\mnras} \textbf{\bibinfo{volume}{416}},
  \bibinfo{pages}{801} (\bibinfo{year}{2011}{\natexlab{a}}),
  \eprint{1102.2903}.

\bibitem[{\citenamefont{{Kay} et~al.}(2012{\natexlab{a}})\citenamefont{{Kay},
  {Peel}, {Short}, {Thomas}, {Young}, {Battye}, {Liddle}, and
  {Pearce}}}]{2012MNRAS.422.1999K}
\bibinfo{author}{\bibfnamefont{S.~T.} \bibnamefont{{Kay}}},
  \bibinfo{author}{\bibfnamefont{M.~W.} \bibnamefont{{Peel}}},
  \bibinfo{author}{\bibfnamefont{C.~J.} \bibnamefont{{Short}}},
  \bibinfo{author}{\bibfnamefont{P.~A.} \bibnamefont{{Thomas}}},
  \bibinfo{author}{\bibfnamefont{O.~E.} \bibnamefont{{Young}}},
  \bibinfo{author}{\bibfnamefont{R.~A.} \bibnamefont{{Battye}}},
  \bibinfo{author}{\bibfnamefont{A.~R.} \bibnamefont{{Liddle}}},
  \bibnamefont{and} \bibinfo{author}{\bibfnamefont{F.~R.}
  \bibnamefont{{Pearce}}}, \bibinfo{journal}{\mnras}
  \textbf{\bibinfo{volume}{422}}, \bibinfo{pages}{1999}
  (\bibinfo{year}{2012}{\natexlab{a}}), \eprint{1112.3769}.

\bibitem[{\citenamefont{{B{\"o}hringer}
  et~al.}(2012)\citenamefont{{B{\"o}hringer}, {Dolag}, and
  {Chon}}}]{2012A&A...539A.120B}
\bibinfo{author}{\bibfnamefont{H.}~\bibnamefont{{B{\"o}hringer}}},
  \bibinfo{author}{\bibfnamefont{K.}~\bibnamefont{{Dolag}}}, \bibnamefont{and}
  \bibinfo{author}{\bibfnamefont{G.}~\bibnamefont{{Chon}}},
  \bibinfo{journal}{\aap} \textbf{\bibinfo{volume}{539}}, \bibinfo{eid}{A120}
  (\bibinfo{year}{2012}), \eprint{1112.5035}.

\bibitem[{\citenamefont{{Kaiser}}(1986)}]{1986MNRAS.222..323K}
\bibinfo{author}{\bibfnamefont{N.}~\bibnamefont{{Kaiser}}},
  \bibinfo{journal}{\mnras} \textbf{\bibinfo{volume}{222}},
  \bibinfo{pages}{323} (\bibinfo{year}{1986}).

\bibitem[{\citenamefont{{Planck Collaboration}
  et~al.}(2011)\citenamefont{{Planck Collaboration}, {Ade}, {Aghanim},
  {Arnaud}, {Ashdown}, {Aumont}, {Baccigalupi}, {Balbi}, {Banday}, {Barreiro}
  et~al.}}]{2011A&A...536A..11P}
\bibinfo{author}{\bibnamefont{{Planck Collaboration}}},
  \bibinfo{author}{\bibfnamefont{P.~A.~R.} \bibnamefont{{Ade}}},
  \bibinfo{author}{\bibfnamefont{N.}~\bibnamefont{{Aghanim}}},
  \bibinfo{author}{\bibfnamefont{M.}~\bibnamefont{{Arnaud}}},
  \bibinfo{author}{\bibfnamefont{M.}~\bibnamefont{{Ashdown}}},
  \bibinfo{author}{\bibfnamefont{J.}~\bibnamefont{{Aumont}}},
  \bibinfo{author}{\bibfnamefont{C.}~\bibnamefont{{Baccigalupi}}},
  \bibinfo{author}{\bibfnamefont{A.}~\bibnamefont{{Balbi}}},
  \bibinfo{author}{\bibfnamefont{A.~J.} \bibnamefont{{Banday}}},
  \bibinfo{author}{\bibfnamefont{R.~B.} \bibnamefont{{Barreiro}}},
  \bibnamefont{et~al.}, \bibinfo{journal}{\aap} \textbf{\bibinfo{volume}{536}},
  \bibinfo{eid}{A11} (\bibinfo{year}{2011}), \eprint{1101.2026}.

\bibitem[{\citenamefont{{Arnaud} et~al.}(2010)\citenamefont{{Arnaud}, {Pratt},
  {Piffaretti}, {B{\"o}hringer}, {Croston}, and
  {Pointecouteau}}}]{2010A&A...517A..92A}
\bibinfo{author}{\bibfnamefont{M.}~\bibnamefont{{Arnaud}}},
  \bibinfo{author}{\bibfnamefont{G.~W.} \bibnamefont{{Pratt}}},
  \bibinfo{author}{\bibfnamefont{R.}~\bibnamefont{{Piffaretti}}},
  \bibinfo{author}{\bibfnamefont{H.}~\bibnamefont{{B{\"o}hringer}}},
  \bibinfo{author}{\bibfnamefont{J.~H.} \bibnamefont{{Croston}}},
  \bibnamefont{and}
  \bibinfo{author}{\bibfnamefont{E.}~\bibnamefont{{Pointecouteau}}},
  \bibinfo{journal}{\aap} \textbf{\bibinfo{volume}{517}}, \bibinfo{eid}{A92}
  (\bibinfo{year}{2010}), \eprint{0910.1234}.

\bibitem[{\citenamefont{{Bora} and {Desai}}(2021)}]{Boraepj}
\bibinfo{author}{\bibfnamefont{K.}~\bibnamefont{{Bora}}} \bibnamefont{and}
  \bibinfo{author}{\bibfnamefont{S.}~\bibnamefont{{Desai}}},
  \bibinfo{journal}{European Physical Journal C} \textbf{\bibinfo{volume}{81}},
  \bibinfo{eid}{296} (\bibinfo{year}{2021}), \eprint{2103.12695}.

\bibitem[{\citenamefont{{Scolnic} et~al.}(2018)\citenamefont{{Scolnic},
  {Jones}, {Rest}, {Pan}, {Chornock}, {Foley}, {Huber}, {Kessler}, {Narayan},
  {Riess} et~al.}}]{pantheon}
\bibinfo{author}{\bibfnamefont{D.~M.} \bibnamefont{{Scolnic}}},
  \bibinfo{author}{\bibfnamefont{D.~O.} \bibnamefont{{Jones}}},
  \bibinfo{author}{\bibfnamefont{A.}~\bibnamefont{{Rest}}},
  \bibinfo{author}{\bibfnamefont{Y.~C.} \bibnamefont{{Pan}}},
  \bibinfo{author}{\bibfnamefont{R.}~\bibnamefont{{Chornock}}},
  \bibinfo{author}{\bibfnamefont{R.~J.} \bibnamefont{{Foley}}},
  \bibinfo{author}{\bibfnamefont{M.~E.} \bibnamefont{{Huber}}},
  \bibinfo{author}{\bibfnamefont{R.}~\bibnamefont{{Kessler}}},
  \bibinfo{author}{\bibfnamefont{G.}~\bibnamefont{{Narayan}}},
  \bibinfo{author}{\bibfnamefont{A.~G.} \bibnamefont{{Riess}}},
  \bibnamefont{et~al.}, \bibinfo{journal}{\apj} \textbf{\bibinfo{volume}{859}},
  \bibinfo{eid}{101} (\bibinfo{year}{2018}), \eprint{1710.00845}.

\bibitem[{\citenamefont{{Riess} et~al.}(2022)\citenamefont{{Riess}, {Yuan},
  {Macri}, {Scolnic}, {Brout}, {Casertano}, {Jones}, {Murakami}, {Anand},
  {Breuval} et~al.}}]{riess22}
\bibinfo{author}{\bibfnamefont{A.~G.} \bibnamefont{{Riess}}},
  \bibinfo{author}{\bibfnamefont{W.}~\bibnamefont{{Yuan}}},
  \bibinfo{author}{\bibfnamefont{L.~M.} \bibnamefont{{Macri}}},
  \bibinfo{author}{\bibfnamefont{D.}~\bibnamefont{{Scolnic}}},
  \bibinfo{author}{\bibfnamefont{D.}~\bibnamefont{{Brout}}},
  \bibinfo{author}{\bibfnamefont{S.}~\bibnamefont{{Casertano}}},
  \bibinfo{author}{\bibfnamefont{D.~O.} \bibnamefont{{Jones}}},
  \bibinfo{author}{\bibfnamefont{Y.}~\bibnamefont{{Murakami}}},
  \bibinfo{author}{\bibfnamefont{G.~S.} \bibnamefont{{Anand}}},
  \bibinfo{author}{\bibfnamefont{L.}~\bibnamefont{{Breuval}}},
  \bibnamefont{et~al.}, \bibinfo{journal}{\apjl}
  \textbf{\bibinfo{volume}{934}}, \bibinfo{eid}{L7} (\bibinfo{year}{2022}),
  \eprint{2112.04510}.

\bibitem[{\citenamefont{{Efstathiou}}(2021)}]{george21}
\bibinfo{author}{\bibfnamefont{G.}~\bibnamefont{{Efstathiou}}},
  \bibinfo{journal}{\mnras} \textbf{\bibinfo{volume}{505}},
  \bibinfo{pages}{3866} (\bibinfo{year}{2021}), \eprint{2103.08723}.

\bibitem[{\citenamefont{{Nunes} and {Di Valentino}}(2021)}]{nunes21}
\bibinfo{author}{\bibfnamefont{R.~C.} \bibnamefont{{Nunes}}} \bibnamefont{and}
  \bibinfo{author}{\bibfnamefont{E.}~\bibnamefont{{Di Valentino}}},
  \bibinfo{journal}{\prd} \textbf{\bibinfo{volume}{104}}, \bibinfo{eid}{063529}
  (\bibinfo{year}{2021}), \eprint{2107.09151}.

\bibitem[{\citenamefont{Camarena and Marra}(2021)}]{Camarena_2021}
\bibinfo{author}{\bibfnamefont{D.}~\bibnamefont{Camarena}} \bibnamefont{and}
  \bibinfo{author}{\bibfnamefont{V.}~\bibnamefont{Marra}},
  \bibinfo{journal}{Monthly Notices of the Royal Astronomical Society}
  \textbf{\bibinfo{volume}{504}}, \bibinfo{pages}{5164–5171}
  (\bibinfo{year}{2021}), ISSN \bibinfo{issn}{1365-2966},
  \urlprefix\url{http://dx.doi.org/10.1093/mnras/stab1200}.

\bibitem[{\citenamefont{Pedregosa et~al.}(2011)\citenamefont{Pedregosa,
  Varoquaux, Gramfort, Michel, Thirion, Grisel, Blondel, Prettenhofer, Weiss,
  Dubourg et~al.}}]{sklearn}
\bibinfo{author}{\bibfnamefont{F.}~\bibnamefont{Pedregosa}},
  \bibinfo{author}{\bibfnamefont{G.}~\bibnamefont{Varoquaux}},
  \bibinfo{author}{\bibfnamefont{A.}~\bibnamefont{Gramfort}},
  \bibinfo{author}{\bibfnamefont{V.}~\bibnamefont{Michel}},
  \bibinfo{author}{\bibfnamefont{B.}~\bibnamefont{Thirion}},
  \bibinfo{author}{\bibfnamefont{O.}~\bibnamefont{Grisel}},
  \bibinfo{author}{\bibfnamefont{M.}~\bibnamefont{Blondel}},
  \bibinfo{author}{\bibfnamefont{P.}~\bibnamefont{Prettenhofer}},
  \bibinfo{author}{\bibfnamefont{R.}~\bibnamefont{Weiss}},
  \bibinfo{author}{\bibfnamefont{V.}~\bibnamefont{Dubourg}},
  \bibnamefont{et~al.}, \bibinfo{journal}{Journal of Machine Learning Research}
  \textbf{\bibinfo{volume}{12}}, \bibinfo{pages}{2825} (\bibinfo{year}{2011}).

\bibitem[{\citenamefont{{Seikel} et~al.}(2012)\citenamefont{{Seikel},
  {Clarkson}, and {Smith}}}]{seikel12}
\bibinfo{author}{\bibfnamefont{M.}~\bibnamefont{{Seikel}}},
  \bibinfo{author}{\bibfnamefont{C.}~\bibnamefont{{Clarkson}}},
  \bibnamefont{and} \bibinfo{author}{\bibfnamefont{M.}~\bibnamefont{{Smith}}},
  \bibinfo{journal}{\jcap} \textbf{\bibinfo{volume}{2012}}, \bibinfo{eid}{036}
  (\bibinfo{year}{2012}), \eprint{1204.2832}.

\bibitem[{\citenamefont{{Benisty} et~al.}(2022)\citenamefont{{Benisty},
  {Mifsud}, {Levi Said}, and {Staicova}}}]{benisty22}
\bibinfo{author}{\bibfnamefont{D.}~\bibnamefont{{Benisty}}},
  \bibinfo{author}{\bibfnamefont{J.}~\bibnamefont{{Mifsud}}},
  \bibinfo{author}{\bibfnamefont{J.}~\bibnamefont{{Levi Said}}},
  \bibnamefont{and}
  \bibinfo{author}{\bibfnamefont{D.}~\bibnamefont{{Staicova}}},
  \bibinfo{journal}{arXiv e-prints} \bibinfo{eid}{arXiv:2202.04677}
  (\bibinfo{year}{2022}), \eprint{2202.04677}.

\bibitem[{\citenamefont{{Singirikonda} and {Desai}}(2020)}]{Haveesh}
\bibinfo{author}{\bibfnamefont{H.}~\bibnamefont{{Singirikonda}}}
  \bibnamefont{and} \bibinfo{author}{\bibfnamefont{S.}~\bibnamefont{{Desai}}},
  \bibinfo{journal}{European Physical Journal C} \textbf{\bibinfo{volume}{80}},
  \bibinfo{eid}{694} (\bibinfo{year}{2020}), \eprint{2003.00494}.

\bibitem[{\citenamefont{{Rozo} et~al.}(2012)\citenamefont{{Rozo}, {Vikhlinin},
  and {More}}}]{More}
\bibinfo{author}{\bibfnamefont{E.}~\bibnamefont{{Rozo}}},
  \bibinfo{author}{\bibfnamefont{A.}~\bibnamefont{{Vikhlinin}}},
  \bibnamefont{and} \bibinfo{author}{\bibfnamefont{S.}~\bibnamefont{{More}}},
  \bibinfo{journal}{\apj} \textbf{\bibinfo{volume}{760}}, \bibinfo{eid}{67}
  (\bibinfo{year}{2012}), \eprint{1202.2150}.

\bibitem[{\citenamefont{{Loken} et~al.}(2002)\citenamefont{{Loken}, {Norman},
  {Nelson}, {Burns}, {Bryan}, and {Motl}}}]{loken02}
\bibinfo{author}{\bibfnamefont{C.}~\bibnamefont{{Loken}}},
  \bibinfo{author}{\bibfnamefont{M.~L.} \bibnamefont{{Norman}}},
  \bibinfo{author}{\bibfnamefont{E.}~\bibnamefont{{Nelson}}},
  \bibinfo{author}{\bibfnamefont{J.}~\bibnamefont{{Burns}}},
  \bibinfo{author}{\bibfnamefont{G.~L.} \bibnamefont{{Bryan}}},
  \bibnamefont{and} \bibinfo{author}{\bibfnamefont{P.}~\bibnamefont{{Motl}}},
  \bibinfo{journal}{\apj} \textbf{\bibinfo{volume}{579}}, \bibinfo{pages}{571}
  (\bibinfo{year}{2002}), \eprint{astro-ph/0207095}.

\bibitem[{\citenamefont{{Galli}}(2013)}]{galli}
\bibinfo{author}{\bibfnamefont{S.}~\bibnamefont{{Galli}}},
  \bibinfo{journal}{\prd} \textbf{\bibinfo{volume}{87}}, \bibinfo{eid}{123516}
  (\bibinfo{year}{2013}), \eprint{1212.1075}.

\bibitem[{\citenamefont{{Cola{\c{c}}o}
  et~al.}(2019)\citenamefont{{Cola{\c{c}}o}, {Holanda}, {Silva}, and
  {Alcaniz}}}]{colaco19}
\bibinfo{author}{\bibfnamefont{L.~R.} \bibnamefont{{Cola{\c{c}}o}}},
  \bibinfo{author}{\bibfnamefont{R.~F.~L.} \bibnamefont{{Holanda}}},
  \bibinfo{author}{\bibfnamefont{R.}~\bibnamefont{{Silva}}}, \bibnamefont{and}
  \bibinfo{author}{\bibfnamefont{J.~S.} \bibnamefont{{Alcaniz}}},
  \bibinfo{journal}{\jcap} \textbf{\bibinfo{volume}{2019}}, \bibinfo{eid}{014}
  (\bibinfo{year}{2019}), \eprint{1901.10947}.

\bibitem[{\citenamefont{{Planelles} et~al.}(2017)\citenamefont{{Planelles},
  {Fabjan}, {Borgani}, {Murante}, {Rasia}, {Biffi}, {Truong},
  {Ragone-Figueroa}, {Granato}, {Dolag} et~al.}}]{planelles17}
\bibinfo{author}{\bibfnamefont{S.}~\bibnamefont{{Planelles}}},
  \bibinfo{author}{\bibfnamefont{D.}~\bibnamefont{{Fabjan}}},
  \bibinfo{author}{\bibfnamefont{S.}~\bibnamefont{{Borgani}}},
  \bibinfo{author}{\bibfnamefont{G.}~\bibnamefont{{Murante}}},
  \bibinfo{author}{\bibfnamefont{E.}~\bibnamefont{{Rasia}}},
  \bibinfo{author}{\bibfnamefont{V.}~\bibnamefont{{Biffi}}},
  \bibinfo{author}{\bibfnamefont{N.}~\bibnamefont{{Truong}}},
  \bibinfo{author}{\bibfnamefont{C.}~\bibnamefont{{Ragone-Figueroa}}},
  \bibinfo{author}{\bibfnamefont{G.~L.} \bibnamefont{{Granato}}},
  \bibinfo{author}{\bibfnamefont{K.}~\bibnamefont{{Dolag}}},
  \bibnamefont{et~al.}, \bibinfo{journal}{\mnras}
  \textbf{\bibinfo{volume}{467}}, \bibinfo{pages}{3827} (\bibinfo{year}{2017}),
  \eprint{1612.07260}.

\bibitem[{\citenamefont{{Biffi} et~al.}(2014)\citenamefont{{Biffi},
  {Sembolini}, {De Petris}, {Valdarnini}, {Yepes}, and
  {Gottl{\"o}ber}}}]{biffi14}
\bibinfo{author}{\bibfnamefont{V.}~\bibnamefont{{Biffi}}},
  \bibinfo{author}{\bibfnamefont{F.}~\bibnamefont{{Sembolini}}},
  \bibinfo{author}{\bibfnamefont{M.}~\bibnamefont{{De Petris}}},
  \bibinfo{author}{\bibfnamefont{R.}~\bibnamefont{{Valdarnini}}},
  \bibinfo{author}{\bibfnamefont{G.}~\bibnamefont{{Yepes}}}, \bibnamefont{and}
  \bibinfo{author}{\bibfnamefont{S.}~\bibnamefont{{Gottl{\"o}ber}}},
  \bibinfo{journal}{\mnras} \textbf{\bibinfo{volume}{439}},
  \bibinfo{pages}{588} (\bibinfo{year}{2014}), \eprint{1401.2992}.

\bibitem[{\citenamefont{{Fabjan}
  et~al.}(2011{\natexlab{b}})\citenamefont{{Fabjan}, {Borgani}, {Rasia},
  {Bonafede}, {Dolag}, {Murante}, and {Tornatore}}}]{fabjan11}
\bibinfo{author}{\bibfnamefont{D.}~\bibnamefont{{Fabjan}}},
  \bibinfo{author}{\bibfnamefont{S.}~\bibnamefont{{Borgani}}},
  \bibinfo{author}{\bibfnamefont{E.}~\bibnamefont{{Rasia}}},
  \bibinfo{author}{\bibfnamefont{A.}~\bibnamefont{{Bonafede}}},
  \bibinfo{author}{\bibfnamefont{K.}~\bibnamefont{{Dolag}}},
  \bibinfo{author}{\bibfnamefont{G.}~\bibnamefont{{Murante}}},
  \bibnamefont{and}
  \bibinfo{author}{\bibfnamefont{L.}~\bibnamefont{{Tornatore}}},
  \bibinfo{journal}{\mnras} \textbf{\bibinfo{volume}{416}},
  \bibinfo{pages}{801} (\bibinfo{year}{2011}{\natexlab{b}}),
  \eprint{1102.2903}.

\bibitem[{\citenamefont{{Kay} et~al.}(2012{\natexlab{b}})\citenamefont{{Kay},
  {Peel}, {Short}, {Thomas}, {Young}, {Battye}, {Liddle}, and
  {Pearce}}}]{kay12}
\bibinfo{author}{\bibfnamefont{S.~T.} \bibnamefont{{Kay}}},
  \bibinfo{author}{\bibfnamefont{M.~W.} \bibnamefont{{Peel}}},
  \bibinfo{author}{\bibfnamefont{C.~J.} \bibnamefont{{Short}}},
  \bibinfo{author}{\bibfnamefont{P.~A.} \bibnamefont{{Thomas}}},
  \bibinfo{author}{\bibfnamefont{O.~E.} \bibnamefont{{Young}}},
  \bibinfo{author}{\bibfnamefont{R.~A.} \bibnamefont{{Battye}}},
  \bibinfo{author}{\bibfnamefont{A.~R.} \bibnamefont{{Liddle}}},
  \bibnamefont{and} \bibinfo{author}{\bibfnamefont{F.~R.}
  \bibnamefont{{Pearce}}}, \bibinfo{journal}{\mnras}
  \textbf{\bibinfo{volume}{422}}, \bibinfo{pages}{1999}
  (\bibinfo{year}{2012}{\natexlab{b}}), \eprint{1112.3769}.

\bibitem[{\citenamefont{{Stanek}
  et~al.}(2010{\natexlab{b}})\citenamefont{{Stanek}, {Rasia}, {Evrard},
  {Pearce}, and {Gazzola}}}]{stanek10}
\bibinfo{author}{\bibfnamefont{R.}~\bibnamefont{{Stanek}}},
  \bibinfo{author}{\bibfnamefont{E.}~\bibnamefont{{Rasia}}},
  \bibinfo{author}{\bibfnamefont{A.~E.} \bibnamefont{{Evrard}}},
  \bibinfo{author}{\bibfnamefont{F.}~\bibnamefont{{Pearce}}}, \bibnamefont{and}
  \bibinfo{author}{\bibfnamefont{L.}~\bibnamefont{{Gazzola}}},
  \bibinfo{journal}{\apj} \textbf{\bibinfo{volume}{715}}, \bibinfo{pages}{1508}
  (\bibinfo{year}{2010}{\natexlab{b}}), \eprint{0910.1599}.

\bibitem[{\citenamefont{{Holanda} and {da Silva}}(2020)}]{2020JCAP...12..027H}
\bibinfo{author}{\bibfnamefont{R.~F.~L.} \bibnamefont{{Holanda}}}
  \bibnamefont{and} \bibinfo{author}{\bibfnamefont{W.~J.~C.} \bibnamefont{{da
  Silva}}}, \bibinfo{journal}{\jcap} \textbf{\bibinfo{volume}{2020}},
  \bibinfo{eid}{027} (\bibinfo{year}{2020}), \eprint{2007.14199}.

\bibitem[{\citenamefont{{Mendon{\c{c}}a}
  et~al.}(2021)\citenamefont{{Mendon{\c{c}}a}, {Bora}, {Holanda}, and
  {Desai}}}]{mendonca21_cddr}
\bibinfo{author}{\bibfnamefont{I.~E.~C.~R.} \bibnamefont{{Mendon{\c{c}}a}}},
  \bibinfo{author}{\bibfnamefont{K.}~\bibnamefont{{Bora}}},
  \bibinfo{author}{\bibfnamefont{R.~F.~L.} \bibnamefont{{Holanda}}},
  \bibnamefont{and} \bibinfo{author}{\bibfnamefont{S.}~\bibnamefont{{Desai}}},
  \bibinfo{journal}{\jcap} \textbf{\bibinfo{volume}{2021}}, \bibinfo{eid}{084}
  (\bibinfo{year}{2021}), \eprint{2107.14169}.

\bibitem[{\citenamefont{{Foreman-Mackey}
  et~al.}(2013)\citenamefont{{Foreman-Mackey}, {Hogg}, {Lang}, and
  {Goodman}}}]{emcee}
\bibinfo{author}{\bibfnamefont{D.}~\bibnamefont{{Foreman-Mackey}}},
  \bibinfo{author}{\bibfnamefont{D.~W.} \bibnamefont{{Hogg}}},
  \bibinfo{author}{\bibfnamefont{D.}~\bibnamefont{{Lang}}}, \bibnamefont{and}
  \bibinfo{author}{\bibfnamefont{J.}~\bibnamefont{{Goodman}}},
  \bibinfo{journal}{\pasp} \textbf{\bibinfo{volume}{125}}, \bibinfo{pages}{306}
  (\bibinfo{year}{2013}), \eprint{1202.3665}.

\bibitem[{\citenamefont{Foreman-Mackey}(2016)}]{corner}
\bibinfo{author}{\bibfnamefont{D.}~\bibnamefont{Foreman-Mackey}},
  \bibinfo{journal}{The Journal of Open Source Software}
  \textbf{\bibinfo{volume}{1}}, \bibinfo{pages}{24} (\bibinfo{year}{2016}),
  \urlprefix\url{https://doi.org/10.21105/joss.00024}.

\bibitem[{\citenamefont{{Limousin} et~al.}(2013)\citenamefont{{Limousin},
  {Morandi}, {Sereno}, {Meneghetti}, {Ettori}, {Bartelmann}, and
  {Verdugo}}}]{2013SSRv..177..155L}
\bibinfo{author}{\bibfnamefont{M.}~\bibnamefont{{Limousin}}},
  \bibinfo{author}{\bibfnamefont{A.}~\bibnamefont{{Morandi}}},
  \bibinfo{author}{\bibfnamefont{M.}~\bibnamefont{{Sereno}}},
  \bibinfo{author}{\bibfnamefont{M.}~\bibnamefont{{Meneghetti}}},
  \bibinfo{author}{\bibfnamefont{S.}~\bibnamefont{{Ettori}}},
  \bibinfo{author}{\bibfnamefont{M.}~\bibnamefont{{Bartelmann}}},
  \bibnamefont{and}
  \bibinfo{author}{\bibfnamefont{T.}~\bibnamefont{{Verdugo}}},
  \bibinfo{journal}{\ssr} \textbf{\bibinfo{volume}{177}}, \bibinfo{pages}{155}
  (\bibinfo{year}{2013}), \eprint{1210.3067}.

\bibitem[{\citenamefont{{Hofmann} et~al.}(2017)\citenamefont{{Hofmann},
  {Sanders}, {Clerc}, {Nandra}, {Ridl}, {Dennerl}, {Ramos-Ceja}, {Finoguenov},
  and {Reiprich}}}]{2017A&A...606A.118H}
\bibinfo{author}{\bibfnamefont{F.}~\bibnamefont{{Hofmann}}},
  \bibinfo{author}{\bibfnamefont{J.~S.} \bibnamefont{{Sanders}}},
  \bibinfo{author}{\bibfnamefont{N.}~\bibnamefont{{Clerc}}},
  \bibinfo{author}{\bibfnamefont{K.}~\bibnamefont{{Nandra}}},
  \bibinfo{author}{\bibfnamefont{J.}~\bibnamefont{{Ridl}}},
  \bibinfo{author}{\bibfnamefont{K.}~\bibnamefont{{Dennerl}}},
  \bibinfo{author}{\bibfnamefont{M.}~\bibnamefont{{Ramos-Ceja}}},
  \bibinfo{author}{\bibfnamefont{A.}~\bibnamefont{{Finoguenov}}},
  \bibnamefont{and} \bibinfo{author}{\bibfnamefont{T.~H.}
  \bibnamefont{{Reiprich}}}, \bibinfo{journal}{\aap}
  \textbf{\bibinfo{volume}{606}}, \bibinfo{eid}{A118} (\bibinfo{year}{2017}),
  \eprint{1708.05205}.

\bibitem[{\citenamefont{{Perotto} et~al.}(2022)\citenamefont{{Perotto}, {Adam},
  {Ade}, {Ajeddig}, {Andr{\'e}}, {Arnaud}, {Artis}, {Aussel}, {Bartalucci},
  {Beelen} et~al.}}]{2022EPJWC.25700038P}
\bibinfo{author}{\bibfnamefont{L.}~\bibnamefont{{Perotto}}},
  \bibinfo{author}{\bibfnamefont{R.}~\bibnamefont{{Adam}}},
  \bibinfo{author}{\bibfnamefont{P.}~\bibnamefont{{Ade}}},
  \bibinfo{author}{\bibfnamefont{H.}~\bibnamefont{{Ajeddig}}},
  \bibinfo{author}{\bibfnamefont{P.}~\bibnamefont{{Andr{\'e}}}},
  \bibinfo{author}{\bibfnamefont{M.}~\bibnamefont{{Arnaud}}},
  \bibinfo{author}{\bibfnamefont{E.}~\bibnamefont{{Artis}}},
  \bibinfo{author}{\bibfnamefont{H.}~\bibnamefont{{Aussel}}},
  \bibinfo{author}{\bibfnamefont{I.}~\bibnamefont{{Bartalucci}}},
  \bibinfo{author}{\bibfnamefont{A.}~\bibnamefont{{Beelen}}},
  \bibnamefont{et~al.}, in \emph{\bibinfo{booktitle}{mm Universe @ NIKA2 -
  Observing the mm Universe with the NIKA2 Camera}} (\bibinfo{year}{2022}),
  vol. \bibinfo{volume}{257} of \emph{\bibinfo{series}{European Physical
  Journal Web of Conferences}}, p. \bibinfo{pages}{00038}, \eprint{2111.01729}.

\bibitem[{\citenamefont{{K{\'e}ruzor{\'e}}
  et~al.}(2022)\citenamefont{{K{\'e}ruzor{\'e}}, {Artis},
  {Mac{\'\i}as-P{\'e}rez}, {Mayet}, {Mu{\~n}oz-Echeverr{\'\i}a}, {Perotto}, and
  {Ruppin}}}]{2022EPJWC.25700025K}
\bibinfo{author}{\bibfnamefont{F.}~\bibnamefont{{K{\'e}ruzor{\'e}}}},
  \bibinfo{author}{\bibfnamefont{E.}~\bibnamefont{{Artis}}},
  \bibinfo{author}{\bibfnamefont{J.~F.} \bibnamefont{{Mac{\'\i}as-P{\'e}rez}}},
  \bibinfo{author}{\bibfnamefont{F.}~\bibnamefont{{Mayet}}},
  \bibinfo{author}{\bibfnamefont{M.}~\bibnamefont{{Mu{\~n}oz-Echeverr{\'\i}a}}},
  \bibinfo{author}{\bibfnamefont{L.}~\bibnamefont{{Perotto}}},
  \bibnamefont{and} \bibinfo{author}{\bibfnamefont{F.}~\bibnamefont{{Ruppin}}},
  in \emph{\bibinfo{booktitle}{mm Universe @ NIKA2 - Observing the mm Universe
  with the NIKA2 Camera}} (\bibinfo{year}{2022}), vol. \bibinfo{volume}{257} of
  \emph{\bibinfo{series}{European Physical Journal Web of Conferences}}, p.
  \bibinfo{pages}{00025}, \eprint{2111.01660}.

\end{thebibliography}

\end{document}